\documentclass[twocolumn,showpacs,preprintnumbers,amsmath,amssymb]{revtex4}
\usepackage{graphicx}% Include figure files
\usepackage{dcolumn}% Align table columns on decimal point
\usepackage{bm}% bold math
\usepackage{latexsym}
\usepackage{amsfonts}
\usepackage{amssymb}
\usepackage{amsmath}
\begin{document}

\preprint{KUNS-1886}

\title{ Braneworld  Kaluza-Klein Corrections in a Nutshell}

\author{Sugumi Kanno}
\email{sugumi@tap.scphys.kyoto-u.ac.jp}
\author{Jiro Soda}
\email{jiro@tap.scphys.kyoto-u.ac.jp}
\affiliation{
 Department of Physics,  Kyoto University, Kyoto 606-8501, Japan
}%

\date{\today}% It is always \today, today,

\begin{abstract}
  Evaluating Kaluza-Klein (KK) corrections is indispensable to test the 
  braneworld scenario. In this paper, we propose  a novel symmetry 
  approach to a 4-dimensional effective action with KK corrections
  for the Randall-Sundrum two-brane system.
  The result can be used to assess the validity of
 the low energy approximation. Also, our result provides the basis
 for predicting CMB spectrum with KK corrections and the study of
 the transition  from black strings to  black holes.
\end{abstract}

\maketitle

%===============================================================%
%************************ SECTION I ****************************%
%===============================================================%

\section{Introduction}

  It is generally believed that the singularity problem of the cosmology
  can be resolved in the context of the superstring theory. It seems
  that the most clear prediction of the superstring theory is the existence of
  extra-dimensions. This apparently contradicts our experience.
  Fortunately, the superstring theory itself provides
  a mechanism to hide extra-dimensions, which is the so-called braneworld 
  scenario where the standard matter lives on the brane, 
  while only the gravity can feel the bulk space-time. 
   This scenario has been realized by Randall and Sundrum
     in a two-brane model~\cite{RS1}. 
  Needless to say, it is important to test this new picture 
  by the cosmological observations in the context of this model. 
  
 We need some efficient calculational method to give quantitative predictions
    to be compared with cosmological observations.
   As the observable quantities are usually represented by the 4-dimensional 
   language, it would be advantageous if we could find purely 4-dimensional
 description of the braneworld which includes the enough information of 
 the bulk  geometry, i.e. KK effects.   
 The purpose of this paper is to derive the 4-dimensional 
 effective action with  KK effects for the two-brane system.  
 For this purpose, we  propose
  a novel symmetry approach that utilize the conformal symmetry as a
 principle  to determine the effective action. 
  Our new method  gives not only a simple re-derivation 
  of known results~\cite{KS1,wiseman},  but also  a new result, 
  i.e. the effective action with KK corrections. 

The organization of this paper is as follows.
 In sec.II, we explain our method and re-derive known results.
 In sec.III, we derive a new result, i.e. the KK corrected effective action.
 In the final section, we summarize our results and discuss
  possible applications and extension of our results. 
 Throughout this paper, we take the unit $8\pi G =1$.

%===============================================================%
%************************ SECTION II ****************************%
%===============================================================%

\section{Symmetry Approach}
 
 In this paper, for simplicity, we concentrate on the vacuum two-brane system. 
 Let us start with the 5-dimensional action  for this system
\begin{eqnarray}
  S [\gamma_{AB}, g_{\mu\nu}, h_{\mu\nu}] 
\end{eqnarray}
where $\gamma_{AB}$, $g_{\mu\nu}$ and $h_{\mu\nu}$ are the 5-dimensional
 bulk metric, the induced metric on the positive and the negative 
 tension branes, respectively. Here, $A,B$ and $\mu, \nu $ 
 label the 5-dimensional and 4-dimensional coordinates, respectively.   
 The variation with respect to $\gamma_{AB}$ gives the bulk Einstein equations
 and the variation with respect to $g_{\mu\nu}$ and $h_{\mu\nu}$ 
 yields the junction  conditions. 
 Now,  suppose to solve the bulk equations of motion and the 
 junction condition on the negative tension brane, 
 then formally we get the relation
\begin{eqnarray}
  \gamma_{AB} = \gamma_{AB}[g_{\mu\nu}] 
  \ , \quad h_{\mu\nu} = h_{\mu\nu} [g_{\mu\nu}] \ .
  \label{bulk-metric}
\end{eqnarray}
By substituting relations (\ref{bulk-metric})  into the original action,
 in principle, the 4-dimensional effective action can be obtained as
\begin{eqnarray}
 S_{\rm eff} 
 = S[\gamma_{AB} [g_{\mu\nu}] ,g_{\mu\nu}, h_{\mu\nu}[g_{\mu\nu}]] \ .
 \label{eff-action}
\end{eqnarray}
 In practice, however, the above calculation  is not feasible. 
 Therefore, in the following, we will give a method 
 to obtain the effective action
 without doing the above calculation. Our method combines the gradient 
 expansion approach and the geometric approach.  

   The gradient expansion approach can be used  at low energy.
 In this case, it is legitimate to assume that 
 the action can be expanded by the local terms with increasing
  orders of derivatives if one includes all of the relevant 
 degrees of freedom~\cite{KS1}. 
 In the two-brane system, the relevant degrees of freedom
 are nothing but the metric  and  the radion  which can be seen from the linear
 analysis~\cite{GT}. 
  Hence, we assume the general local action constructed from the 
  metric $g_{\mu\nu}$ and the radion $\Psi $ as an ansatz. 
 Therefore,  we can write the action as 
\begin{eqnarray}
S_{\rm eff}
\!\!&=&\!\!
	{1\over 2} \int d^4 x \sqrt{-g} \left[ \Psi R - 2\Lambda (\Psi )
	-{\omega (\Psi) \over \Psi} \nabla^\mu \Psi \nabla_\mu \Psi \right]
        \nonumber\\
&&\!\!
	+\int d^4 x \sqrt{-g} \left[
	A(\Psi) \left( \nabla^\mu \Psi \nabla_\mu \Psi \right)^2
	+B(\Psi) \left( \Box \Psi \right)^2 \right. \nonumber \\
&&\!\!\!\!\left.
	+C(\Psi)  \nabla^\mu \Psi \nabla_\mu \Psi \Box \Psi
	+D(\Psi) R~\Box \Psi \right. \nonumber\\
&&\left.\!\!\!\!  
	+ E(\Psi) R \nabla^\mu \Psi \nabla_\mu \Psi
     + F(\Psi) R^{\mu\nu} \nabla_\mu \Psi \nabla_\nu \Psi 
     + G(\Psi) R^2     \right. \nonumber\\
 && \left.\!\!\!\!
	+ H(\Psi) R^{\mu\nu} R_{\mu\nu} 
	+I(\Psi) R^{\mu\nu\lambda\rho} R_{\mu\nu\lambda\rho} 
	+\cdots  \right]  \ , 
	\label{setup}
\end{eqnarray}
where $\nabla_\mu$ denotes the covariant derivative with respect to the metric
$g_{\mu\nu}$. Here, we have listed up all of the possible local terms 
which have derivatives up to fourth-order. This series will continue infinitely.
This can be regarded as  the generalization of the scalar-tensor theory
  including the higher derivative terms. 
 We have the freedom to redefine the scalar field $\Psi$. In fact, we have used
 this freedom to fix the functional form of the coefficient of $R$. 
 However, we can not determine other coefficient functions without any 
 information about the bulk geometry. 

 The geometric approach  yields,  instead of the action,  directly 
 the effective equations of motion~\cite{ShiMaSa} 
\begin{eqnarray}
  G_{\mu\nu} = T_{\mu\nu} + \pi_{\mu\nu} - E_{\mu\nu}
  \label{SMS}  
\end{eqnarray}
where $T_{\mu\nu}$ is the energy-momentum tensor of the matter and
\begin{eqnarray}
  \pi_{\mu\nu} = -{1\over 4} T_{\mu\lambda} T^\lambda{}_{\nu}
      + {1\over 12} T T_{\mu\nu} 
    +{1\over 8}g_{\mu\nu} \left( T^{\alpha\beta} T_{\alpha\beta} 
      -{1\over 3} T^2 \right) \ . 
\end{eqnarray}
Note that the projection of Weyl tensor $E_{\mu\nu}$ represents  the
 effect of the bulk geometry. 
For the vacuum brane, we can put 
$T_{\mu\nu} + \pi_{\mu\nu} = - \lambda g_{\mu\nu}$. Hence, 
 Eq.~(\ref{SMS}) reduces to 
\begin{eqnarray}
   G_{\mu\nu} =  - E_{\mu\nu} - \lambda g_{\mu\nu}   \ .
\end{eqnarray}
One defect of this approach is that $E_{\mu\nu}$ is not determined
 except for the following property
\begin{eqnarray}
   E^\mu{}_\mu =0 \ .
   \label{traceless}
\end{eqnarray}
 For the isotropic homogeneous universe, Eq.~(\ref{traceless}) 
 has sufficient information to
 deduce the cosmological evolution equation
\begin{eqnarray}
    H^2 = {\lambda \over 3} + {C \over a^4} \ ,
\end{eqnarray}
where $C$ is the constant of integration. This effect of the bulk acts as 
 radiation fluid, hence it is called as dark radiation. 
  For general spacetimes, however, this traceless condition is not sufficient
  to determine the evolution of the braneworld. 
 
 In the former approach, we have introduced the radion explicitly. While
 the radion never appears in the latter approach, instead  $E_{\mu\nu}$ 
 is induced as the effective energy-momentum tensor reflecting the
 effects of the bulk geometry.  
 Notice that the  property (\ref{traceless}) implies the conformal invariance 
 of this effective matter. Clearly, both approaches should agree to each other.
  Hence, the radion  must play a role of the conformally invariant matter 
 $E_{\mu\nu}$.   This requirement gives a stringent constraint on the
 action, more precisely,  the conformal symmetry (\ref{traceless}) determines
 radion dependent  coefficients  in the action (\ref{setup}). 

 Let us illustrate our method using the following truncated action 
\begin{eqnarray}
    S_{\rm eff} 
    ={1\over 2}  \int d^4 x \sqrt{-g} \left[ 
          \Psi R -2\Lambda (\Psi) - {\omega (\Psi) \over \Psi} 
          \nabla^\mu \Psi \nabla_\mu \Psi \right] \ ,
\end{eqnarray}
 which is nothing but the scalar-tensor theory with coupling function
 $\omega (\Psi)$ and the potential function $\Lambda (\Psi )$. 
 Note that this is the most general local action which contains 
 up to the second  order derivatives and has the general coordinate invariance.
 It should be stressed that the scalar-tensor theory is, in general,
  not related to the braneworld. However, we know a special type of 
 scalar-tensor theory corresponds to  the low energy 
 braneworld~\cite{KS1,wiseman}. 
 Here, we will present a simple derivation of this known fact.  
 First, we must find $E_{\mu\nu}$. 
The above action gives the equations of motion for the metric as
\begin{eqnarray}
    G_{\mu\nu} &=& -{\Lambda \over \Psi} g_{\mu\nu}
                   + {1\over \Psi} \left( 
                 \nabla_\mu \nabla_\nu \Psi - g_{\mu\nu} \Box \Psi \right)
                 \nonumber\\
      && \quad            + {\omega \over \Psi^2} \left(
             \nabla_\mu \Psi \nabla_\nu \Psi  -{1\over 2} g_{\mu\nu}
             \nabla^\alpha \Psi \nabla_\alpha \Psi \right)  \ . 
\end{eqnarray}
The right hand side of this Eq.~(11) should be identified with 
$-E_{\mu\nu}-\lambda g_{\mu\nu}$.
 Hence, the  condition  $E^\mu{}_\mu =0$ becomes
\begin{eqnarray}
        \Box \Psi = - {\omega \over 3\Psi} 
        \nabla^\mu \Psi \nabla_\mu \Psi  
        - {4\over 3} \left( \Lambda  - \lambda \Psi \right)  \ .
\end{eqnarray}
This is the equation for the radion $\Psi$. However, we also
have the equation for $\Psi$ from the action as 
\begin{eqnarray}
    \Box \Psi = \left( {1\over 2\Psi} - { \omega' \over 2\omega} \right)
             \nabla^\alpha \Psi \nabla_\alpha \Psi  
             - {\Psi \over 2\omega} R + {\Psi \over \omega} \Lambda'    \ ,
\end{eqnarray}
where the prime denotes the derivative with respect to $\Psi$. 
In order for these two Eqs.~(12) and (13) to be compatible, $\Lambda$ and
  $\omega$ must satisfy 
\begin{eqnarray}
&&-{\omega \over 3 \Psi} = {1\over 2\Psi} - { \omega' \over 2\omega}  \ ,  \\
&&{4\over 3} \left( \Lambda - \lambda \Psi \right) =
            {\Psi \over \omega} \left( 2\lambda - \Lambda' \right)  \ ,
\end{eqnarray}
where we used  $R= 4\lambda$ which comes from the trace part of Eq.~(7).
 Eqs.~(14) and (15) can be integrated as  
\begin{eqnarray}
   \Lambda (\Psi) = \lambda + \lambda \beta \left( 1-\Psi \right)^2   \ , \quad
  \omega (\Psi ) = {3\over 2} {\Psi \over 1-\Psi}  \ ,  
\end{eqnarray}
where the constant of integration $\beta$ represents the ratio
 of the cosmological constant on the negative tension brane to that on 
 the positive tension brane. 
 Here, one of constants of integration is absorbed by rescaling of $\Psi$.
 In doing so, we have assumed the constant of integration is positive.
 We can also describe the negative tension brane if we take the 
 negative signature.

Thus, we get the effective action 
\begin{eqnarray}
S_{\rm eff}
	&=&\int d^4 x \sqrt{-g} \left[ {1\over 2} \Psi R 
	-{3 \over 4( 1-\Psi )} \nabla^\mu \Psi \nabla_\mu \Psi \right. 
	\nonumber\\
	&&\left.  \qquad \qquad \qquad
    - \lambda - \lambda \beta (1-\Psi)^2 \right] \ .
\end{eqnarray}
Surprisingly, this completely agrees with the previous 
result~\cite{KS1,wiseman}. Surprinsingly enough, our simple symmetry 
principle $E^\mu{}_\mu =0$ has determined the action completely. 
 
 As we have shown in \cite{KSS}, if $\beta <-1$
 there exists a static deSitter two-brane solution
 which turns out to be unstable. In particular,  
 two inflating branes  can collide at $\Psi = 0$. 
 This process is completely smooth for the observer on the brane.  
 This fact led us to the born-again scenario.
 The similar process occurs also in the ekpyrotic (cyclic) model~\cite{turok}
 where the moduli approximation is used. It can be shown that the moduli
 approximation is nothing but the lowest order truncation of the low energy
 gradient expansion method developed by us~\cite{KS1}. Hence,
 it is of great interest to see the leading order corrections due
 to KK modes to this process.  

%===============================================================%
%************************ SECTION III ***************************%
%===============================================================%

\section{KK corrections}

  Let us extend the result in the previous section
 to the higher order case. 
 We have already determined the functions $\Lambda (\Psi)$ and $\omega (\Psi )$.
 From the linear analysis, the action in the previous section is known 
 to come only from zero modes. Hence, one can expect the other coefficients 
 in the action (\ref{setup}) represent the effects of KK-modes. 
 In the case of the single-brane model, this contribution
 is non-local. While, in the two-brane model, as we have shown explicitly
 in ~\cite{KS1}, 
 the KK-effects can be represented by the local terms if we introduce
 the radion field. 
 
  Now we impose the conformal symmetry  on the fourth order derivative terms
  in the action (\ref{setup}) as we did in the  previous section. 
  Starting from the action (\ref{setup}), one can read off the equation for 
  the metric
 and hence $E_{\mu\nu}$ can be identified. 
 The compatibility condition between $E^\mu{}_\mu =0$ and the equation for the
 radion $\Psi$ leads to
\begin{eqnarray}
  &&  (1-\Psi) (C'' -3A') = C' + 3E''+{3\over 2} F'' \\
  &&  (1-\Psi) (2B'' -4A) = 2B' +C + 3D'' + 3E'+{5\over 2} F' \qquad \\
  &&  (1-\Psi) (4C' -8A) = 2C + 12 E' + 5F' \\
  &&  (1-\Psi) (3B'-2C) = 2B + 3 D' + F   \\ 
  &&  4(1-\Psi) B' = 2B + 6 D' + 6E + 3F 
\end{eqnarray}
\begin{eqnarray}
  &&  (1-\Psi) C =  3E +F   \\
  && 2 (1-\Psi) B = 3 D \\
  &&  (1-\Psi) (2C - F') = 6E + 2F + 2H'' + 4I'' \\
  &&  (1-\Psi) F = - H' - 2 I' \\
  && (1-\Psi ) (D'' - E' ) = D' + 6 G'' + H''  \\
  &&  2 (1-\Psi ) (D' -E ) = D + 6 G' + H'   \\
  &&  G' = H' = I' = 0  \ .
\end{eqnarray}
These equations seem to be over constrained. Nevertheless, one can find
 solutions consistently. 
From Eqs.~(26) and (29), we see $F=0$. 
Eqs.~(23) and (24) can be solved as
\begin{eqnarray}
 E = {1\over 3} (1-\Psi ) C \ , \quad 
   D =  {2\over 3} (1-\Psi ) B \ . 
\end{eqnarray}
Substituting these results into Eqs.~(18) -(22) and Eqs.~(27) and (28), we have
\begin{eqnarray}
  && 3 (1-\Psi ) A' = C'  \ , \\ 
  && (1-\Psi ) (C' + 4A ) = 2B' \ ,\\
  && 4(1-\Psi ) A = C  \ , \\
  && B'-2C = 0   \ , \\
  &&  (1-\Psi ) C = B \ , \\
  &&  2B = (1-\Psi ) \left[ 
      6B' - 2(1-\Psi ) B'' -C + (1-\Psi ) C' \right] \ , \qquad \\
  && 3B = (1- \Psi ) (2B' -C)   \ .
\end{eqnarray}
Combining  Eqs.~(34) and (35), we obtain
\begin{eqnarray}
      B = {\ell^2 \over (1-\Psi )^2 }  \ , 
\end{eqnarray}
where $\ell^2$ is the constant of integration representing the curvature
 scale of the bulk.  
 Eqs.~(35) and (33) give 
\begin{eqnarray}
    C= {\ell^2 \over (1-\Psi )^3} \ , \quad 
    A= {1\over 4}{\ell^2 \over (1-\Psi)^4} \ .
\end{eqnarray}
The rest of Eqs.~(31), (32), (36) and (37) are identically satisfied. 
 The coefficients $G, H$ and $I$ must be constants $g,h$ and $i$.  
 Because of the existence of the Gauss-Bonnet topological term, we can put $i=0$
 without losing the generality. The constants $g$ and $h$ can be interpreted as 
 the variety of the effects of the bulk gravitational waves.
 
Thus, we find the 4-dimensional effective action with KK corrections as
\begin{eqnarray}
S_{\rm eff} 
  &=& \int d^4 x \sqrt{-g} \left[ {1\over 2} \Psi R 
     - {3 \over 4( 1-\Psi )} \nabla^\mu \Psi \nabla_\mu \Psi \right. \nonumber\\
   &&  \left.  
    - \lambda - \lambda \beta (1-\Psi)^2    \right]
              \nonumber\\
 &&    + \ell^2 \int d^4 x \sqrt{-g} \left[
     {1 \over 4 (1-\Psi)^4} \left( \nabla^\mu \Psi \nabla_\mu \Psi \right)^2
                            \right. \nonumber \\
 && \left.    + {1\over (1-\Psi)^2} \left( \Box \Psi \right)^2 
      + {1\over (1-\Psi)^3} \nabla^\mu \Psi \nabla_\mu \Psi\Box \Psi
      \right. \nonumber \\
 && \left.   + {2\over 3(1-\Psi)} R \Box \Psi 
      + {1\over 3(1-\Psi)^2}  R \nabla^\mu \Psi \nabla_\mu \Psi 
      \right. \nonumber \\
 && \left.     + g R^2     + h R^{\mu\nu} R_{\mu\nu}  \right] \ .
 \label{action}
\end{eqnarray}
It should be noted that this action becomes non-local after
 integrating out the radion field. This fits the fact that
 KK effects are non-local usually.  
  In principle, we can continue this calculation to any order of 
 derivatives. 
 
 There are many possible applications using our effective action
 which will be published in separate papers~\cite{Sugumi}. 
 Here, we only mention the collision of two branes at low energy.
 In the ekpyrotic (cyclic) model and born-again model, the cosmological
 evolution was completely smooth at the collision time in the Jordan frame
 which is physical in the braneworld picture. Here, we can say this is true
 even if we take into account KK corrections. This is seen from
 the fact the action (\ref{action}) is regular at the collision point 
 $\Psi =0$.

%===============================================================%
%************************ SECTION IV ****************************%
%===============================================================%

\section{Conclusion}

  We have established  a novel symmetry 
  approach to  a 4-dimensional effective action with KK corrections. 
  This is done by combining the low energy expansion of the action
  and the geometric approach. Our result supports the smoothness
  of the collision process of two branes advocated 
  in the ekpyrotic (cyclic) model and  born-again model. 
  Our result can be used to assess the validity of
 the low energy approximation. It also has a potential to make concrete
 predictions to be compared with observations. 

 As to the cosmological applications, it is important to recognize
 that our action can describe the inflation.  
 Cosmological perturbations~\cite{soda} are
 now ready to be studied. 
 In fact, our result provides the basis
 of the prediction of CMB spectrum with KK corrections.
 
 The black hole solutions with KK corrections
 are also interesting subjects. 
 If we truncate the system at the lowest order, the static solution 
 is Schwartzshild black hole with a trivial radion which corresponds 
 to the black string in the bulk. 
 The Gregorry-Laflamme instability occurs when the wavelength of KK modes
 exceeds the gravitational length of the black hole~\cite{GL}. 
 Clearly, the lightest
 KK mode is important and this mode is already included in our action, 
 hence it would be  interesting to investigate if 
  the Gregory-Laflamme instability occurs or not within our 
  theory~\cite{KS2}.

  As an extension of our analysis, we can incorporate the matter fields
  both on the brane and in the bulk. 
  As to the matter on the brane, 
  at the lowest order, we obtain the same functional for $\omega$
  and the coupling the radion with the matter is that 
  previously obtained~\cite{KS1}. 
  Inclusion of KK-effects is also possible if we take the scalar field
  as the matter~\cite{Sugumi}. It is rather straightforward to incorporate 
  bulk fields into our scheme.

\begin{acknowledgements}
 The authors would like to thank Kei-ichi Maeda for useful comments. 
This work was supported in part by  Grant-in-Aid for  Scientific
Research Fund of the Ministry of Education, Science and Culture of Japan 
 No. 155476 (SK) and  No.14540258 (JS) and also
  by a Grant-in-Aid for the 21st Century COE ``Center for
  Diversity and Universality in Physics".  
\end{acknowledgements}


\begin{thebibliography}{99}

\bibitem{RS1}
L.~Randall and R.~Sundrum,
%``A large mass hierarchy from a small extra dimension,''
Phys.\ Rev.\ Lett.\  {\bf 83}, 3370 (1999)
[arXiv:hep-ph/9905221].
%%CITATION = HEP-PH 9905221;%%


\bibitem{KS1}
S.~Kanno and J.~Soda,
%``Radion and holographic brane gravity,''
Phys.\ Rev.\ D {\bf 66}, 083506 (2002)
[arXiv:hep-th/0207029];
%%CITATION = HEP-TH 0207029;%%
S.~Kanno and J.~Soda,
%``Brane world effective action at low energies and AdS/CFT,''
Phys.\ Rev.\ D {\bf 66}, 043526 (2002)
[arXiv:hep-th/0205188].
%%CITATION = HEP-TH 0205188;%%

\bibitem{wiseman}
T.~Chiba,
%``Scalar-tensor gravity in two 3-brane system,''
Phys.\ Rev.\ D {\bf 62}, 021502 (2000)
[arXiv:gr-qc/0001029];
%%CITATION = GR-QC 0001029;%%
T.~Wiseman,
%``Strong brane gravity and the radion at low energies,''
Class.\ Quant.\ Grav.\  {\bf 19}, 3083 (2002)
[arXiv:hep-th/0201127];
%%CITATION = HEP-TH 0201127;%%
%
P.~Brax, C.~van de Bruck, A.~C.~Davis and C.~S.~Rhodes,
%``Cosmological evolution of brane world moduli,''
arXiv:hep-th/0209158;
%%CITATION = HEP-TH 0209158;%%
T.~Shiromizu and K.~Koyama,
%``Low energy effective theory for two brane systems: Covariant curvature  formulation,''
arXiv:hep-th/0210066.
%%CITATION = HEP-TH 0210066;%%



\bibitem{GT}
J.~Garriga and T.~Tanaka,
%``Gravity in the brane-world,''
Phys.\ Rev.\ Lett.\  {\bf 84}, 2778 (2000)
[arXiv:hep-th/9911055].
%%CITATION = HEP-TH 9911055;%%


\bibitem{ShiMaSa}
T.~Shiromizu, K.~Maeda and M.~Sasaki,
%``The Einstein equations on the 3-brane world,''
Phys.\ Rev.\ D {\bf 62}, 024012 (2000)
[arXiv:gr-qc/9910076].
%%CITATION = GR-QC 9910076;%%
 
\bibitem{KSS}
S.~Kanno, M.~Sasaki and J.~Soda,
%``Born-again braneworld,''
Prog. Theor. Phys. {\bf 109}, 357 (2003), 
arXiv:hep-th/0210250.
%%CITATION = HEP-TH 0210250;%%

\bibitem{turok}
J.~Khoury, B.~A.~Ovrut, P.~J.~Steinhardt and N.~Turok,
%``The ekpyrotic universe: Colliding branes and the origin 
%of the hot big  bang,''
Phys.\ Rev.\ D {\bf 64}, 123522 (2001)
[arXiv:hep-th/0103239];
%%CITATION = HEP-TH 0103239;%%
%
%
P.~J.~Steinhardt and N.~Turok,
%``A Cyclic Model Of The Universe,''
Science {\bf 296}, 1436 (2002).
%%CITATION = SCIEA,296,1436;%%

\bibitem{Sugumi}
S.~Kanno and J.Soda, in preparation.
\bibitem{soda}
D.~Langlois,
%``Evolution of cosmological perturbations in a brane-universe,''
Phys.\ Rev.\ Lett.\  {\bf 86}, 2212 (2001)
[arXiv:hep-th/0010063];
%%CITATION = HEP-TH 0010063;%%
H.~Kodama, A.~Ishibashi and O.~Seto,
%``Brane world cosmology: Gauge-invariant formalism for perturbation,''
Phys.\ Rev.\ D {\bf 62}, 064022 (2000)
[arXiv:hep-th/0004160];
%%CITATION = HEP-TH 0004160;%%
C.~van de Bruck, M.~Dorca, R.~H.~Brandenberger and A.~Lukas,
%``Cosmological perturbations in brane-world theories: Formalism,''
Phys.\ Rev.\ D {\bf 62}, 123515 (2000)
[arXiv:hep-th/0005032];
%%CITATION = HEP-TH 0005032;%%
K.~Koyama and J.~Soda,
%``Evolution of cosmological perturbations in the brane world,''
Phys.\ Rev.\ D {\bf 62}, 123502 (2000)
[arXiv:hep-th/0005239];
%%CITATION = HEP-TH 0005239;%%
D.~Langlois, R.~Maartens, M.~Sasaki and D.~Wands,
%``Large-scale cosmological perturbations on the brane,''
Phys.\ Rev.\ D {\bf 63}, 084009 (2001)
[arXiv:hep-th/0012044];
%%CITATION = HEP-TH 0012044;%%
K.~Koyama and J.~Soda,
%``Bulk gravitational field and cosmological perturbations on the brane,''
Phys.\ Rev.\ D {\bf 65}, 023514 (2002)
[arXiv:hep-th/0108003].
%%CITATION = HEP-TH 0108003;%%


\bibitem{GL}
R.~Gregory and R.~Laflamme,
%``Black Strings And P-Branes Are Unstable,''
Phys.\ Rev.\ Lett.\  {\bf 70}, 2837 (1993)
[arXiv:hep-th/9301052].
%%CITATION = HEP-TH 9301052;%%
\bibitem{KS2}
S.~Kanno and J.~Soda,
%``Rotating black string and effective Teukolsky equation in braneworld,''
arXiv:gr-qc/0311074.
%%CITATION = GR-QC 0311074;%%





\end{thebibliography}

\end{document}